\documentclass[aps,prx,twocolumn,secnumarabic,amsmath,amssymb,superscriptaddress]{revtex4-1}
\usepackage{amssymb,amsbsy,graphicx,subfigure,bm,multirow,epstopdf}
\usepackage{color}

\begin{document}
\title{Observing spin fractionalization in the Kitaev spin liquid via temperature evolution of indirect resonant inelastic x-ray scattering}

\author{G\'abor B. Hal\'asz}
\affiliation{Materials Science and Technology Division, Oak Ridge
National Laboratory, Oak Ridge, TN 37831, USA} \affiliation{Kavli
Institute for Theoretical Physics, University of California, Santa
Barbara, CA 93106, USA}
\author{Stefanos Kourtis}
\affiliation{Department of Physics, Boston University, Boston, MA
02215, USA}
\author{Johannes Knolle}
\affiliation{Blackett Laboratory, Imperial College London, London
SW7 2AZ, United Kingdom}
\author{Natalia B. Perkins}
\affiliation{School of Physics and Astronomy, University of
Minnesota, Minneapolis, MN 55455, USA}

\begin{abstract}

Motivated by the ongoing effort to search for high-resolution
signatures of quantum spin liquids, we investigate the temperature
dependence of the indirect resonant inelastic x-ray scattering
(RIXS) response for the Kitaev honeycomb model. We find that, as a
result of spin fractionalization, the RIXS response changes
qualitatively at two well-separated temperature scales, $T_L$ and
$T_H$, which correspond to the characteristic energies of the two
kinds of fractionalized excitations, $\mathbb{Z}_2$ gauge fluxes and
Majorana fermions, respectively. While thermally excited
$\mathbb{Z}_2$ gauge fluxes at temperature $T_L$ lead to a general
broadening and softening of the response, the thermal proliferation
of Majorana fermions at temperature $T_H \sim 10 \, T_L$ results in
a significant shift of the spectral weight, both in terms of energy
and momentum. Due to its exclusively indirect nature, the RIXS
process we consider gives rise to a universal magnetic response and,
from an experimental perspective, it directly corresponds to the
$K$-edge of Ru$^{3+}$ in the Kitaev candidate material
$\alpha$-RuCl$_3$.

\end{abstract}

\maketitle

\pagebreak

\section{Introduction}

Recent years have seen tremendous interest in Kitaev materials
\cite{Kitaev2006, Jackeli2009, Chaloupka2010, Cao2013, Witczak2014,
Rau2016, Trebst2017, Hermanns2017, Winter2017}, a family of
spin-orbit-assisted Mott insulators on tri-coordinated
two-dimensional (2D) and three-dimensional (3D) lattices, in which
local, spin-orbit-entangled $j_{\rm eff}=1/2$ moments interact via
strongly bond-directional Ising-like interactions. The most
extensively studied Kitaev materials are the iridates A$_2$IrO$_3$
(A = Li, Na) \cite{Singh2010, Liu2011, Singh2012, Ye2012, Comin2012,
Biffin2014a, Biffin2014b, Takayama2015, Chun2015, Williams2016} and
H$_3$LiIr$_2$O$_6$ \cite{Kitagawa2018}, and the ruthenium compound
$\alpha$-RuCl$_3$ \cite{Plumb2014, Sears2015, Majumder2015,
Johnson2015}. The interest in these materials originates from the
belief that they are proximate to the Kitaev quantum spin liquid
(QSL) \cite{Kitaev2006} due to the presence of dominant Kitaev
interactions in their microscopic Hamiltonians \cite{Chun2015,
Rau2014, Yamaji2014, Sizyuk2014, Rousochatzakis2015, Winter2016,
Katukuri2016, Yadav2016}.

When searching for QSL physics in Kitaev materials, a general
feature to look for is the fractionalization of spins into two types
of quasiparticle excitations, according to the exact solution of the
Kitaev model \cite{Kitaev2006}: localized, gapped $\mathbb{Z}_2$
fluxes and itinerant, gapless Majorana fermions. In pursuit of spin
fractionalization, a lot of experimental and theoretical effort has
been devoted to the study of spin dynamics in Kitaev materials
through various dynamical probes, such as inelastic neutron
scattering (INS) \cite{Choi2012, Knolle2014a, Knolle2015, Smith2015,
Banerjee2016, Smith2016, Song2016, Banerjee2017, Do2017,
Knolle2018}, Raman scattering \cite{Knolle2014b, Sandilands2015,
Perreault2015, Sandilands2016, Glamazda2016, Perreault2016a,
Perreault2016b, Wang2018}, and resonant inelastic x-ray scattering
(RIXS) \cite{Halasz2016, Halasz2017}. The key idea is that, even if
residual magnetic order sets in below a critical temperature, which
indeed happens in most of the Kitaev materials, the fractionalized
quasiparticles of the nearby QSL phase may still lead to observable
signatures in the dynamical response \cite{Banerjee2016,
Rousochatzakis2018}.

\begin{figure}[!t]
\includegraphics[width=0.9\columnwidth]{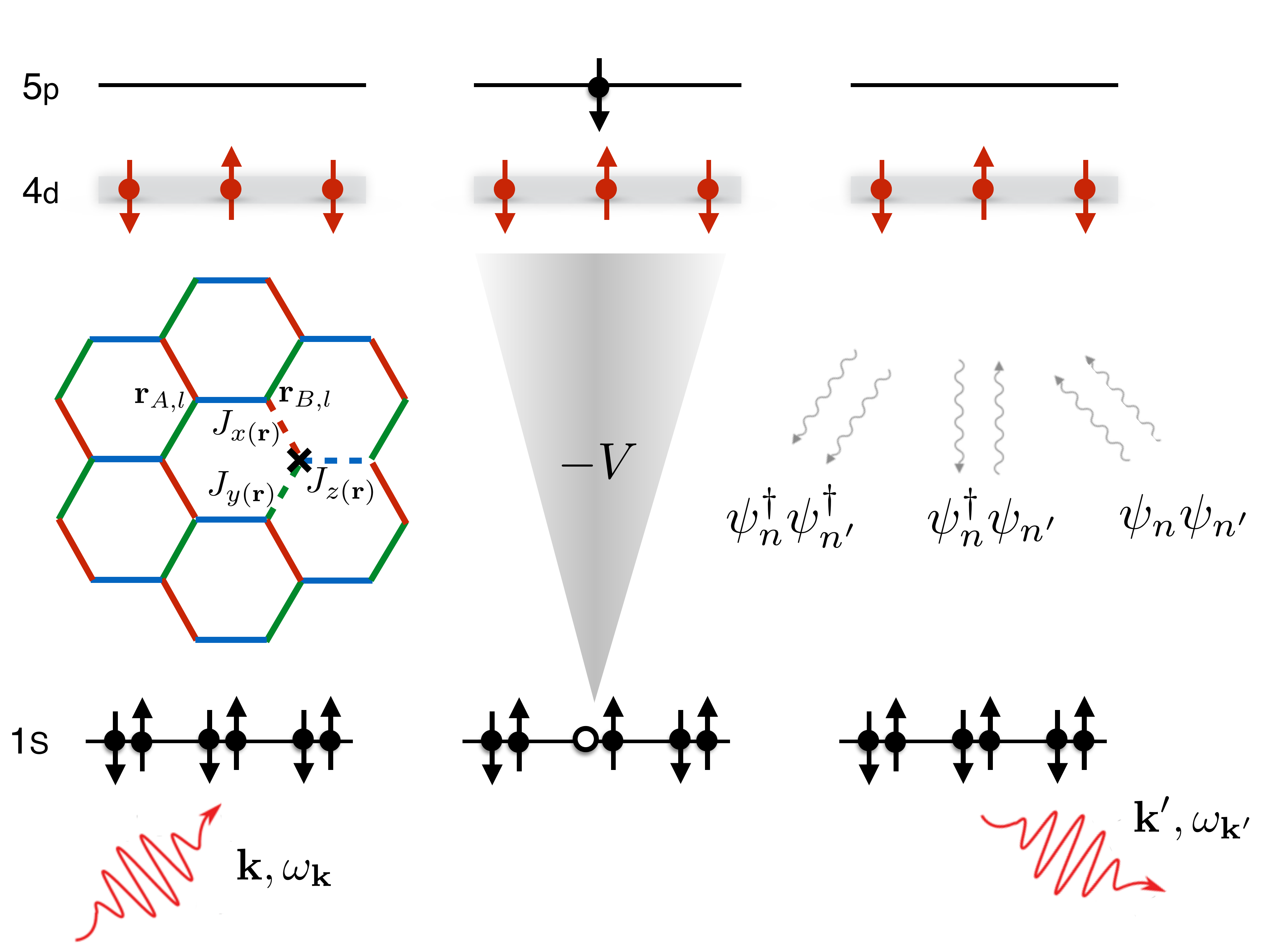}
\caption{Illustration of an indirect RIXS process at the Ru$^{3+}$
$K$-edge creating Majorana fermion excitations in the Kitaev
honeycomb model, due to the local modification of magnetic couplings
in the intermediate state. The three bond types $x$, $y$, and $z$ of
the lattice are marked by red, green, and blue, respectively, while
the bonds with modified couplings $J_{\kappa (\mathbf{r})}$ between
the photon-scattering site $\mathbf{r}$ and the neighboring sites
$\kappa (\mathbf{r})$ (with $\kappa = x,y,z$) are denoted by dashed
lines. The incoming (outgoing) x-ray photons have momenta
$\mathbf{k}$ ($\mathbf{k}'$) and energies $\omega_{\mathbf{k}} = c
|\mathbf{k}|$ ($\omega_{\mathbf{k}'} = c |\mathbf{k}'|$). At finite
temperature, three kinds of indirect RIXS processes contribute to
the response: Stokes processes creating two fermions, anti-Stokes
processes annihilating two fermions, and ``mixed'' processes
creating one fermion while annihilating another one.}\label{fig-1}
\end{figure}

In particular, there is a growing body of experimental evidence that
the ground state of the spin-orbit-assisted honeycomb Mott insulator
$\alpha$-RuCl$_3$ is proximate to the Kitaev QSL phase, despite the
fact that it exhibits zigzag antiferromagnetic order below $T_N
\simeq 7$ K \cite{Plumb2014, Sears2015, Johnson2015, Majumder2015}.
For example, the INS response of $\alpha$-RuCl$_3$
\cite{Banerjee2016, Banerjee2017, Do2017} shows a broad continuum
spectrum of 2D magnetic fluctuations around the center of the
Brillouin zone, which is indicative of spin fractionalization and is
in agreement with the corresponding theoretical prediction for the
Kitaev honeycomb model \cite{Knolle2014a, Knolle2015}. Promising
results were also obtained by Raman scattering experiments in
$\alpha$-RuCl$_3$, detecting  a broad continuum below 100 K that
even persists into the magnetically ordered phase
\cite{Sandilands2015, Sandilands2016}. Moreover, the temperature
dependence of the Raman spectral weight can be interpreted in terms
of the spins fractionalizing into fermionic quasiparticles
\cite{Wang2018}.

As a general spectroscopic probe of magnetic materials, RIXS has
important advantages over both INS and Raman scattering. In contrast
to INS, which only measures dynamic single-spin correlations, and
Raman scattering, which is restricted to essentially zero momentum
due to its low-energy photons, RIXS offers greater versatility in
measuring a wider range of dynamic correlations with full momentum
resolution \cite{Ament2011a, Ament2011b, Savary2015, Kim2017}.
Specifically, for the Kitaev QSL, it was predicted by some of us
that the magnetic channels of RIXS are capable of picking up both
types of fractionalized excitations \cite{Halasz2016, Halasz2017}.
Indeed, while the non-spin-conserving channels are dominated by the
localized $\mathbb{Z}_2$ fluxes and thus give rise to a weakly
dispersive response, the spin-conserving channel couples exclusively
to the Majorana fermions and can effectively probe the
characteristic graphene-like dispersion of these exotic
fractionalized quasiparticles.

Nevertheless, spin fractionalization in the Kitaev materials has not
yet been observed in RIXS experiments due to at least two major
difficulties in designing a suitable measurement. First, in order to
distinguish between the various magnetic channels, one would need to
do polarization analysis on the outgoing x-ray beam. Second, the
energy resolution of RIXS at the previously proposed $L_3$-edge
\cite{Halasz2016, Halasz2017} is rather poor, both in the iridates
and in $\alpha$-RuCl$_3$. In this work, we instead propose that
signatures of fractionalized excitations in $\alpha$-RuCl$_3$ can be
probed by \emph{indirect} RIXS at the $K$-edge of Ru$^{3+}$. In
addition to a favorable predicted energy resolution \cite{Gog2013,
Gog2016}, this edge has only one magnetic channel due to its
indirect nature, and the corresponding magnetic response is thus
independent of x-ray polarization.

Furthermore, it is now well appreciated that, due to the flat band
of low-energy $\mathbb{Z}_2$ fluxes, the dynamical responses of the
Kitaev QSL are rather sensitive to thermal fluctuations. Indeed,
already at temperatures corresponding to only a small fraction of
the Kitaev exchange energy, thermal population of the fluxes
\cite{Nasu2015} can give rise to finite-temperature responses that
are strikingly different from their zero-temperature counterparts
\cite{Nasu2016, Yoshitake2016, Yoshitake2017a, Yoshitake2017b}. To
provide a useful guide for experimentalists, we therefore calculate
the indirect RIXS response of the Kitaev QSL at \emph{finite
temperature} and describe how the temperature evolution of this
response reflects the spin-fractionalization scheme in the Kitaev
QSL.

Our main result is that there are qualitative changes in the RIXS
response at two distinct temperature scales, $T_L$ and $T_H$,
separated by an order of magnitude, which correspond to the
characteristic energies of the $\mathbb{Z}_2$ fluxes and the
Majorana fermions, respectively. At the scale of $T_L$, the fluxes
become thermally excited and give rise to an effective disorder for
the Majorana fermions, thereby leading to an overall
\emph{broadening} of the response as well as the softening of the
quasi-sharp features present at zero temperature. At the scale of
$T_H \sim 10 \, T_L$, the Majorana fermions become excited in large
numbers, leading to an overall \emph{shift} of the spectral weight,
both from positive to negative energies and from the boundary to the
center of the Brillouin zone. In the high-temperature regime, we
also identify a pronounced peak in the spectral weight around zero
energy and momentum, corresponding to collective energy-density
fluctuations, and we argue that this peak is related to the
quasi-elastic peak in the experimental Raman response of
$\alpha$-RuCl$_3$ \cite{Sandilands2015, Wang2018}.

\section{Indirect RIXS in the Kitaev quantum spin liquid}

In RIXS experiments, core electrons of a specific ion are promoted
to an unoccupied state using an x-ray beam, thereby locally exciting
the irradiated material into a highly energetic and very short-lived
($\sim 1$ fs) intermediate state. Motivated by $\alpha$-RuCl$_3$, we
are interested in RIXS processes at the $K$-edge of Ru$^{3+}$ (see
Fig.~\ref{fig-1}), which involve the excitation of an electron from
the $1s$ core shell into an unoccupied $5p$ state \emph{above} the
$4d$ valence shell. Since no electrons are excited directly into the
valence orbitals, magnetic excitations can only be created by
\emph{indirect} RIXS processes, which do not change the spin of the
valence shell and thus correspond to the spin-conserving channel
discussed in Refs.~\onlinecite{Halasz2016} and
\onlinecite{Halasz2017}. Consequently, the $K$-edge of Ru$^{3+}$ has
only one magnetic RIXS channel, giving rise to a universal magnetic
response that does not depend on the x-ray polarizations.

In $\alpha$-RuCl$_3$, the magnetism of each Ru$^{3+}$ ion is
governed by a $j_{\mathrm{eff}} = 1/2$ Kramers doublet in the
$t_{2g}$ orbitals of the $4d$ valence shell \cite{Jackeli2009}, and
we assume that the effective low-energy Hamiltonian acting on these
Kramers doublets is that of the Kitaev model \cite{Kitaev2006}:
\begin{equation}
H = -J \sum_{\langle \mathbf{r}, \mathbf{r}' \rangle_x}
\sigma_{\mathbf{r}}^x \sigma_{\mathbf{r}'}^x - J \sum_{\langle
\mathbf{r}, \mathbf{r}' \rangle_y} \sigma_{\mathbf{r}}^y
\sigma_{\mathbf{r}'}^y - J \sum_{\langle \mathbf{r}, \mathbf{r}'
\rangle_z} \sigma_{\mathbf{r}}^z \sigma_{\mathbf{r}'}^z,
\label{eq-H-1}
\end{equation}
where the three bond types $\kappa = x,y,z$ are distinct in their
orientations (see Fig.~\ref{fig-1}). Using the Kitaev fermionization
$\sigma_{\mathbf{r}}^{\kappa} = i b_{\mathbf{r}}^{\kappa}
c_{\mathbf{r}}^{\phantom{\kappa}}$, this Hamiltonian can be written
as $H = J \sum_{\kappa} \sum_{\langle \mathbf{r}, \mathbf{r}'
\rangle_{\kappa}} i u_{\mathbf{r}, \mathbf{r}'}^{\kappa}
c_{\mathbf{r}}^{\phantom{\kappa}}
c_{\mathbf{r}'}^{\phantom{\kappa}}$ in terms of the Majorana
fermions $b_{\mathbf{r}}^{\kappa}$ and
$c_{\mathbf{r}}^{\phantom{\kappa}}$, where $u_{\mathbf{r},
\mathbf{r}'}^{\kappa} \equiv i b_{\mathbf{r}}^{\kappa}
b_{\mathbf{r}'}^{\kappa}$. Importantly, $u_{\mathbf{r},
\mathbf{r}'}^{\kappa}$ are commuting constants of motion, and they
give rise to static flux degrees of freedom $\Pi_{\langle
\mathbf{r}, \mathbf{r}' \rangle \in p} u_{\mathbf{r},
\mathbf{r}'}^{\kappa} = \pm 1$ at the plaquettes $p$. Moreover, in
each flux sector characterized by $u_{\mathbf{r},
\mathbf{r}'}^{\kappa} = \pm 1$, one obtains a free-fermion
Hamiltonian for $c_{\mathbf{r}}^{\phantom{\kappa}}$, which can thus
be identified as deconfined Majorana-fermion excitations. Since
there are no flux excitations in the ground state, it belongs to the
flux sector with $\Pi_{\langle \mathbf{r}, \mathbf{r}' \rangle \in
p} u_{\mathbf{r}, \mathbf{r}'}^{\kappa} = +1$ for all $p$.

During the RIXS process, a momentum $\mathbf{q} = \mathbf{k} -
\mathbf{k}'$ and an energy $\omega = \omega_{\mathbf{k}} -
\omega_{\mathbf{k}'} = c \, \{ |\mathbf{k}| - |\mathbf{k}'| \}$ is
transferred into the Kitaev QSL, where $\mathbf{k}$ and
$\mathbf{k}'$ are the momenta of the incoming and the outgoing x-ray
photons, respectively (see Fig.~\ref{fig-1}). In the intermediate
state, the $1s$ core hole acts like a nonmagnetic impurity and
locally modifies (i.e., strengthens or weakens) the coupling
strength $J$ of the effective Kitaev model \cite{Brink2007}. The
intermediate state is then an eigenstate $| \tilde{n}_{\mathbf{r}}
\rangle$ of the perturbed Kitaev model
\begin{equation}
\tilde{H}_{\mathbf{r}} = H - \delta J \sum_{\kappa = x,y,z}
\sigma_{\mathbf{r}}^{\kappa} \sigma_{\kappa (\mathbf{r})}^{\kappa},
\label{eq-H-2}
\end{equation}
where $\kappa (\mathbf{r})$ is the site connected to the core-hole
site $\mathbf{r}$ by a $\kappa$ bond. Note that the change in the
Kitaev coupling strength, $\delta J$, may be positive or negative
and that the limit of a nonmagnetic vacancy \cite{Willans2010,
Willans2011, Halasz2014}, corresponding to, for example, the
$L$-edges of Ru$^{3+}$ or Ir$^{4+}$ \cite{Halasz2016, Halasz2017},
is recovered by setting $\delta J = -J$.

Due to the indirect nature of the RIXS process considered, the
Kramers-Heisenberg formula \cite{Ament2011a} for the RIXS vertex
takes the simplified form
\begin{equation}
R (\mathbf{q}) = \sum_{\mathbf{r}} e^{i \mathbf{q} \cdot \mathbf{r}}
\sum_{\tilde{n}_{\mathbf{r}}} \frac{| \tilde{n}_{\mathbf{r}} \rangle
\langle \tilde{n}_{\mathbf{r}} |} {\Omega - E_{\tilde{n}} + i
\Gamma} \, , \label{eq-R-1}
\end{equation}
where $\Gamma$ is the core-hole decay rate, $E_{\tilde{n}}$ is the
energy of the intermediate state $| \tilde{n}_{\mathbf{r}} \rangle$,
and $\Omega$ is the energy of the incoming x-ray photon with respect
to the $K$-edge resonance energy. Considering the experimentally
relevant fast-collision regime, where $J \ll \Gamma$, we may assume
that resonance is close enough, such that $\Omega \ll \Gamma$, and
expand Eq.~(\ref{eq-R-1}) in $(\Omega - E_{\tilde{n}}) / \Gamma$ up
to the order of $1 / \Gamma^2$. Exploiting
$\sum_{\tilde{n}_{\mathbf{r}}} | \tilde{n}_{\mathbf{r}} \rangle
\langle \tilde{n}_{\mathbf{r}} | = 1$ as well as
$\sum_{\tilde{n}_{\mathbf{r}}} E_{\tilde{n}} |
\tilde{n}_{\mathbf{r}} \rangle \langle \tilde{n}_{\mathbf{r}} | =
\tilde{H}_{\mathbf{r}}$, and neglecting any terms giving rise to
exclusively elastic responses, the lowest-order RIXS vertex then
becomes
\begin{equation}
R (\mathbf{q}) = \frac{\delta J} {\Gamma^2} \sum_{\mathbf{r}} e^{i
\mathbf{q} \cdot \mathbf{r}} \sum_{\kappa}
\sigma_{\mathbf{r}}^{\kappa} \sigma_{\kappa (\mathbf{r})}^{\kappa}.
\label{eq-R-2}
\end{equation}
This result has a straightforward physical interpretation: the
indirect RIXS vertex in Eq.~(\ref{eq-R-2}) is due to additional
exchange interactions of strength $\delta J$ that are temporarily
switched on around the core-hole site $\mathbf{r}$ in the
short-lived (lifetime: $\tau \sim 1 / \Gamma$) intermediate state
\footnote{The first $1 / \Gamma$ factor in Eq.~(\ref{eq-R-2}) comes
from the broadening of the resonant transition due to the short
core-hole lifetime [see the denominator of Eq.~(\ref{eq-R-1})] and
is generic to all RIXS processes in the fast-collision regime. The
second $1 / \Gamma$ factor comes from the time evolution in the
short-lived intermediate state and is generic to all \emph{indirect}
RIXS processes in the fast-collision regime}.

\section{Finite-temperature response}

The main result of this work is the calculation of the RIXS response
at finite temperature, which first requires a finite-temperature
formulation of the underlying Kitaev model \cite{Nasu2015, Nasu2016,
Yoshitake2016, Yoshitake2017a, Yoshitake2017b}. Qualitatively,
thermal spin fractionalization in the Kitaev model manifests itself
in successive entropy releases at two well-separated temperature
scales $T_L$ and $T_H$. At low temperatures ($T \ll T_L$), the
fluxes are completely frozen and only a small number of Majorana
fermions are thermally excited. At intermediate temperatures ($T_L
\lesssim T \lesssim T_H$), thermal energy goes into both fluxes and
Majorana fermions, but their fractionalized nature remains readily
observable. Finally, at high temperatures ($T \gg T_H$), fluxes and
Majorana fermions recombine into spins, and the system crosses over
to a conventional paramagnetic regime.

Instead of a full and numerically costly Monte Carlo sampling of
flux excitations \cite{Nasu2015}, a quantitative approximation of
the finite-temperature behavior is obtained by taking a random
average over ``typical'' flux sectors and solving the free-fermion
problem in each flux sector exactly \cite{Metavitsiadis2017}. Each
``typical'' flux sector at temperature $T$ is obtained by creating
two flux excitations around each bond with probability $P_T$ such
that the resulting probability of a flux excitation at any plaquette
is
\begin{equation}
\frac{1 - (1 - 2 P_T)^6} {2} = f_T (\Delta) \equiv \frac{1} {1 +
\exp (\Delta / T)} \, , \label{eq-P-1}
\end{equation}
where $\Delta \approx 0.15J$ is the single-flux gap. The solution
for this probability is given by
\begin{equation}
P_T = \frac{1 - [1 - 2 f_T (\Delta)]^{1/6}} {2} \, . \label{eq-P-2}
\end{equation}
In each ``typical'' flux sector, the free-fermion Hamiltonian then
takes the form
\begin{equation}
\mathcal{H} = J \sum_{\kappa} \sum_{\langle \mathbf{r}, \mathbf{r}'
\rangle_{\kappa}} i \bar{u}_{\mathbf{r}, \mathbf{r}'} c_{\mathbf{r}}
c_{\mathbf{r}'}, \label{eq-H-3}
\end{equation}
where each $\bar{u}_{\mathbf{r}, \mathbf{r}'} \equiv \langle
u_{\mathbf{r}, \mathbf{r}'}^{\kappa} \rangle$ is $+1$ with
probability $1 - P_T$ and $-1$ with probability $P_T$.

Exploiting the bipartite nature of the honeycomb lattice, and noting
that each unit cell $l$ has two sites $\mathbf{r}_{A,l}$ and
$\mathbf{r}_{B,l}$ in the two sublattices $A$ and $B$, this
free-fermion Hamiltonian can be written as $\mathcal{H} =
\sum_{l,l'} i M_{ll'} c_{A,l} c_{B,l'}$, where $M_{ll'} = J
\bar{u}_{\mathbf{r}_{A,l}, \mathbf{r}_{B,l'}}$ if $\mathbf{r}_{A,l}$
and $\mathbf{r}_{B,l'}$ are connected and $M_{ll'} = 0$ otherwise.
Note also that $c_{A,l} \equiv c_{\mathbf{r}_{A,l}}$ and $c_{B,l}
\equiv c_{\mathbf{r}_{B,l}}$. Finally, the free-fermion Hamiltonian
is recast into the canonical form $\mathcal{H} = \sum_n
\varepsilon_n (\psi_n^{\dag} \psi_n^{\phantom{\dag}} - 1/2)$, where
the fermions $\psi_n = (\gamma_{A,n} + i \gamma_{B,n}) / 2$, in
terms of $\gamma_{A,n} = \sum_l U_{ln} c_{A,l}$ and $\gamma_{B,n} =
\sum_l V_{ln} c_{B,l}$, and their energies $\varepsilon_n = 2
\Lambda_{nn}$ are obtained from the singular-value decomposition $M
= U \cdot \Lambda \cdot V^T$ \footnote{Since $M$ is a real matrix,
$U$ and $V$ are real orthogonal matrices, while $\Lambda$ is a
diagonal matrix with non-negative (real) entries.}.

In any given flux sector, the lowest-order RIXS vertex in
Eq.~(\ref{eq-R-2}) can be expressed in terms of the fermions as
\begin{eqnarray}
\mathcal{R} (\mathbf{q}) &=& -\frac{\delta J} {J \Gamma^2}
\sum_{l,l'} i M_{ll'} c_{A,l} c_{B,l'} \left( e^{i \mathbf{q} \cdot
\mathbf{r}_{A,l}} + e^{i \mathbf{q} \cdot \mathbf{r}_{B,l'}}
\right) \nonumber \\
&\propto& \sum_{n,n'} \big( \psi_n^{\phantom{\dag}} + \psi_n^{\dag}
\big) \big( \psi_{n'}^{\phantom{\dag}} - \psi_{n'}^{\dag} \big)
\label{eq-R-3} \\
&& \qquad \times \left[ W_A (\mathbf{q}) \cdot \Lambda + \Lambda
\cdot W_B (\mathbf{q}) \right]_{nn'}, \nonumber
\end{eqnarray}
where we introduce $[S_{A/B} (\mathbf{q})]_{ll'} \equiv \delta_{ll'}
e^{i \mathbf{q} \cdot \mathbf{r}_{A/B,l}}$ as well as $W_A
(\mathbf{q}) \equiv U^T \cdot S_A (\mathbf{q}) \cdot U$ and $W_B
(\mathbf{q}) \equiv V^T \cdot S_B (\mathbf{q}) \cdot V$. Finally, by
neglecting all elastic terms that do not change any fermion numbers
and separating inelastic terms that change fermion numbers in
inequivalent ways, the RIXS vertex in Eq.~(\ref{eq-R-3}) can be
written as
\begin{equation}
\mathcal{R} (\mathbf{q}) \propto \sum_{n < n'} \left[
\mathcal{R}^{(1)}_{nn'} (\mathbf{q}) + \mathcal{R}^{(2)}_{nn'}
(\mathbf{q}) \right] + \sum_{n \neq n'} \mathcal{R}^{(3)}_{nn'}
(\mathbf{q}). \label{eq-R-4}
\end{equation}
In particular, the first term describes Stokes processes creating
two fermions each:
\begin{equation}
\mathcal{R}^{(1)}_{nn'} (\mathbf{q}) = -\psi_n^{\dag}
\psi_{n'}^{\dag} [\mathcal{A}_{-} (\mathbf{q})]_{nn'},
\label{eq-R-4-1}
\end{equation}
the second term describes anti-Stokes processes annihilating two
fermions each:
\begin{equation}
\mathcal{R}^{(2)}_{nn'} (\mathbf{q}) = \psi_n^{\phantom{\dag}}
\psi_{n'}^{\phantom{\dag}} [\mathcal{A}_{-} (\mathbf{q})]_{nn'},
\label{eq-R-4-2}
\end{equation}
and the third term describes ``mixed'' processes creating one
fermion and annihilating one fermion each:
\begin{equation}
\mathcal{R}^{(3)}_{nn'} (\mathbf{q}) = \psi_n^{\dag}
\psi_{n'}^{\phantom{\dag}} [\mathcal{A}_{+} (\mathbf{q})]_{nn'},
\label{eq-R-4-3}
\end{equation}
where $\mathcal{A}_{\pm} (\mathbf{q}) = \{W_A (\mathbf{q}) \pm W_B
(\mathbf{q}), \Lambda\}_{\pm}$ in terms of the (anti)commutator
$\{a,b\}_{\pm} \equiv a \cdot b \pm b \cdot a$. \\

\begin{figure*}
\includegraphics[width=0.96\textwidth]{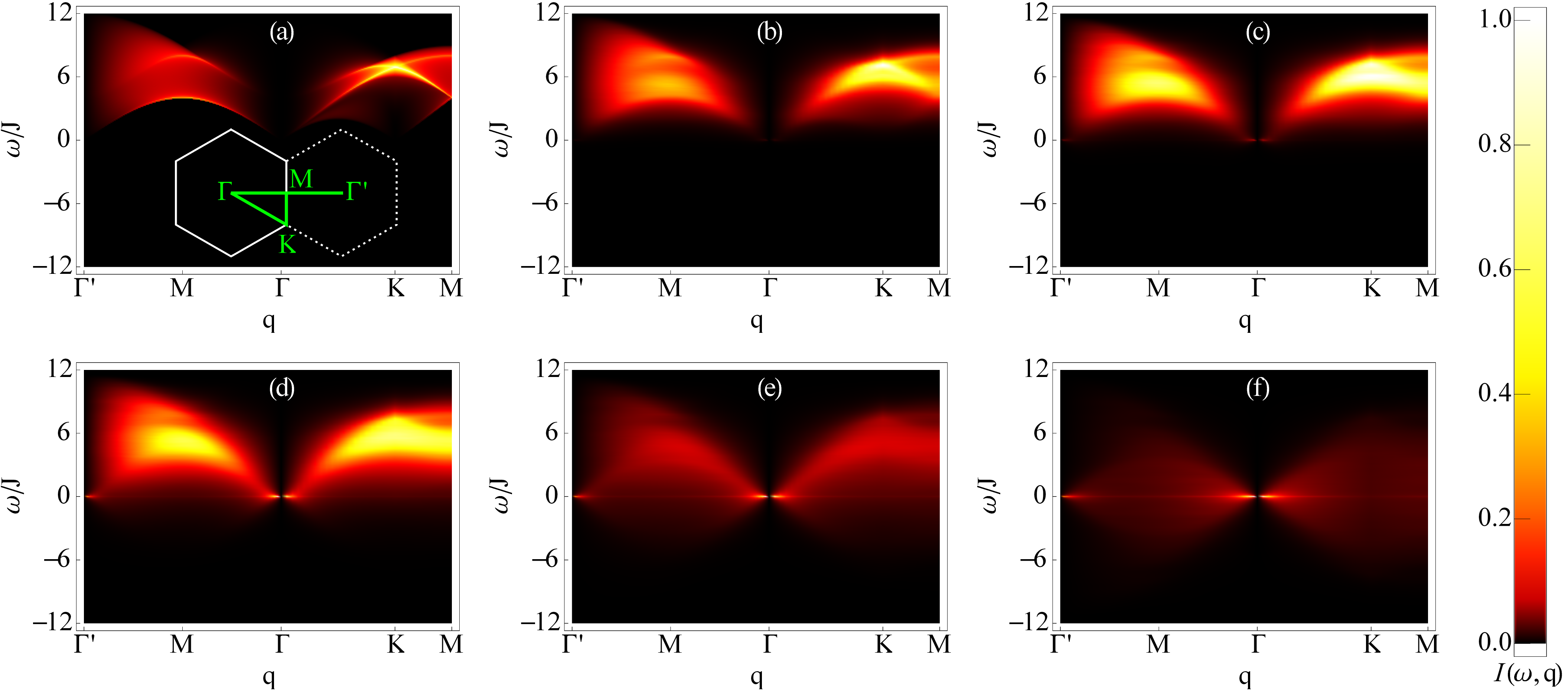}
\caption{Temperature evolution of the indirect RIXS response for the
Kitaev honeycomb model. The lowest-order RIXS intensity $I (\omega,
\mathbf{q})$ is plotted for temperatures (a) $T = 0$, (b) $T =
0.1J$, (c) $T = 0.2J$, (d) $T = 0.5J$, (e) $T = J$, and (f) $T = 5J$
along the high-symmetry path $\Gamma^{\prime}$-M-$\Gamma$-K-M in the
Brillouin zone [marked by green line in the inset of subfigure
(a)].} \label{fig-2}
\end{figure*}

At finite temperature $T$, the resulting RIXS intensity of any given
flux sector is given by Fermi's golden rule:
\begin{equation}
\mathcal{I} (\omega, \mathbf{q}) = \sum_{m,m'} \frac{e^{-E_m / T}}
{Z} \left| \langle m' | \mathcal{R} (\mathbf{q}) | m \rangle
\right|^2 \delta (\omega + E_m - E_{m'}), \label{eq-I-1}
\end{equation}
where $Z \equiv \sum_m e^{-E_m / T}$ is the partition function.
Importantly, the free-fermion eigenstates $| m \rangle = \prod_n
(\psi_n^{\dag})^{N_n} | 0 \rangle$ with energies $E_m = \sum_n N_n
\varepsilon_n$ are labeled by the fermion occupation numbers $N_n =
\{ 0,1 \}$. Since the various terms in Eq.~(\ref{eq-R-4}) all change
the fermion numbers in inequivalent ways, there can be no
interference between them in Eq.~(\ref{eq-I-1}) and their
corresponding intensities can be calculated independently. Moreover,
since the fermions do not interact, the matrix elements of the terms
$\mathcal{R}^{(1,2,3)}_{nn'} (\mathbf{q})$ only depend on the
fermions $n$ and $n'$ whose numbers they actually change.
Substituting Eq.~(\ref{eq-R-4}) into Eq.~(\ref{eq-I-1}), the
lowest-order RIXS intensity is then
\begin{widetext}
\begin{eqnarray}
\mathcal{I} (\omega, \mathbf{q}) &\propto& \, \mathcal{I}^{(1)}
(\omega, \mathbf{q}) + \mathcal{I}^{(2)} (\omega, \mathbf{q}) +
\mathcal{I}^{(3)} (\omega, \mathbf{q}), \nonumber \\
\nonumber \\
\mathcal{I}^{(1)} (\omega, \mathbf{q}) &=& \sum_{n < n'} \left[ 1 -
f_T (\varepsilon_n) \right] \left[ 1 - f_T (\varepsilon_{n'})
\right] \left| \mathcal{A}_{-} (\mathbf{q}) \right|_{nn'}^2 \delta
(\omega - \varepsilon_n - \varepsilon_{n'}),
\nonumber \\
\mathcal{I}^{(2)} (\omega, \mathbf{q}) &=& \sum_{n < n'} f_T
(\varepsilon_n) f_T (\varepsilon_{n'}) \left| \mathcal{A}_{-}
(\mathbf{q}) \right|_{nn'}^2 \delta (\omega + \varepsilon_n +
\varepsilon_{n'}),
\label{eq-I-2} \\
\mathcal{I}^{(3)} (\omega, \mathbf{q}) &=& \sum_{n \neq n'} \left[ 1
- f_T (\varepsilon_n) \right] f_T (\varepsilon_{n'}) \left|
\mathcal{A}_{+} (\mathbf{q}) \right|_{nn'}^2 \delta (\omega -
\varepsilon_n + \varepsilon_{n'}), \nonumber
\end{eqnarray}
\end{widetext}
where the three distinct terms correspond to Stokes, anti-Stokes,
and ``mixed'' processes, respectively. In the limit of $T
\rightarrow 0$, the Fermi functions $f_T (\varepsilon_n)$ vanish for
$\varepsilon_n > 0$, implying that only Stokes processes are
allowed.

In principle, the finite-temperature RIXS intensity of the Kitaev
model is obtained by taking an average of the intensities
corresponding to randomly selected ``typical'' flux sectors: $I
(\omega, \mathbf{q}) = \overline{\mathcal{I} (\omega, \mathbf{q})}$.
For large enough system sizes, however, there are no observable
differences between the intensities of the individual flux sectors.
In practice, it is therefore sufficient to approximate the average
intensity with the intensity corresponding to \emph{any} ``typical''
flux sector: $I (\omega, \mathbf{q}) = \mathcal{I} (\omega,
\mathbf{q})$.

\section{Results and discussion}

\begin{figure}
\includegraphics[width=0.99\columnwidth]{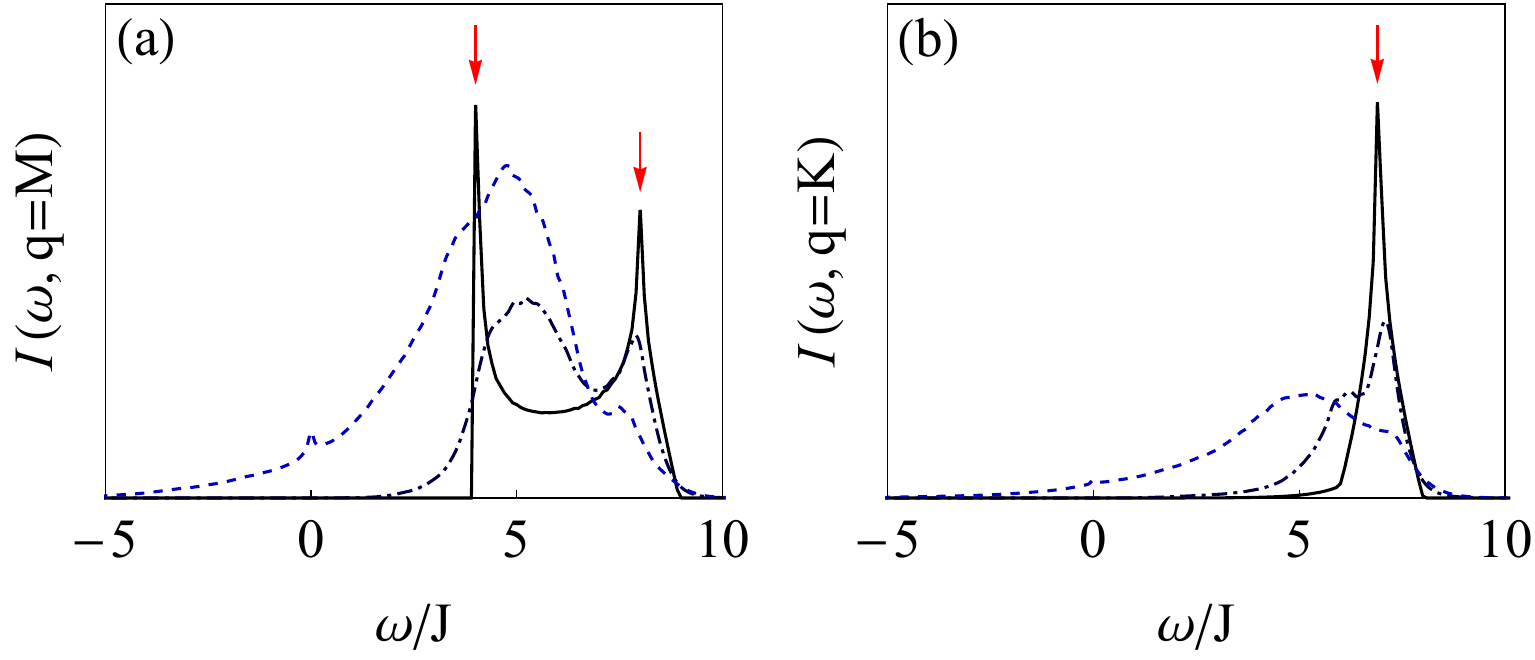}
\caption{Lowest-order RIXS intensities $I (\omega, \mathbf{q})$ at
the M point (a) and at the K point (b) of the Brillouin zone for
temperatures $T = 0$ (solid line), $T = 0.1J$ (dashed-dotted line),
and $T = J$ (dashed line). Red arrows indicate quasi-sharp features
(i.e., logarithmic divergences) at $T = 0$.}\label{fig-3}
\end{figure}

The lowest-order RIXS intensity $I (\omega, \mathbf{q})$ is plotted
in Figs.~\ref{fig-2} and \ref{fig-3} for a range of different
temperatures $T$, along an entire high-symmetry path and at specific
high-symmetry points of the Brillouin zone, respectively. We start
by briefly discussing the limit of zero temperature
\cite{Halasz2016}, in which case the fermions $\psi_n \equiv
\psi_{\mathbf{k}}$ are labeled by their momenta $\mathbf{k}$, and
the matrix element $|\mathcal{A}_{-} (\mathbf{q})|_{\mathbf{k},
\mathbf{k}'}$ in Eq.~(\ref{eq-I-2}) vanishes unless $\mathbf{q} =
\mathbf{k} + \mathbf{k}'$. The RIXS intensity $I (\omega,
\mathbf{q}) \propto \sum_{\mathbf{k}} |\mathcal{A}_{-}
(\mathbf{q})|_{\mathbf{k}, \mathbf{q} - \mathbf{k}}^2 \, \delta
(\omega - \varepsilon_{\mathbf{k}} - \varepsilon_{\mathbf{q} -
\mathbf{k}})$ can then be understood in terms of the characteristic
momentum dispersion $\varepsilon_{\mathbf{k}}$ of the fermions
\cite{Halasz2016, Halasz2017}.

Ignoring the matrix element $[\mathcal{A}_{-}
(\mathbf{q})]_{\mathbf{k}, \mathbf{q} - \mathbf{k}}$, the RIXS
intensity at each momentum $\mathbf{q}$ is proportional to the joint
density of states $\hat{g}_{\omega} (\mathbf{q}) = \sum_{\mathbf{k}}
\delta (\omega - \varepsilon_{\mathbf{k}} - \varepsilon_{\mathbf{q}
- \mathbf{k}})$, which in turn corresponds to an effective (joint)
band dispersion $\hat{\varepsilon}_{\mathbf{k}} (\mathbf{q}) \equiv
\varepsilon_{\mathbf{k}} + \varepsilon_{\mathbf{q} - \mathbf{k}}$ as
a function of the fermion momentum $\mathbf{k}$. Due to the finite
width of this effective band, the RIXS intensity is nonzero for a
finite energy range at each momentum $\mathbf{q}$, which can be
identified as an indirect signature of fractionalization
\cite{Halasz2016}. However, while truly sharp features $I (\omega,
\mathbf{q}) \propto \delta (\omega - \hat{\omega}_{\mathbf{q}})$ are
absent from the RIXS response, there are clear quasi-sharp features
$I (\omega, \mathbf{q}) \propto -\log (\omega -
\hat{\omega}_{\mathbf{q}})$ [see Fig.~\ref{fig-3}] due to
logarithmic divergences in $\hat{g}_{\omega} (\mathbf{q})$, which
correspond to van Hove singularities of the effective band
$\hat{\varepsilon}_{\mathbf{k}} (\mathbf{q})$.

As the temperature is increased, there are qualitative changes in
the RIXS response at the two characteristic temperature scales $T_L
\approx \Delta$ and $T_H \approx J$, which can be identified as
indirect signatures of the flux and fermion excitations,
respectively. At temperatures $T \gtrsim T_L$, thermally excited
fluxes behave like disorder from the perspective of the fermions
\cite{Metavitsiadis2017}, and fermion momentum is therefore no
longer a good quantum number. As a result of this effective
disorder, the quasi-sharp features of the zero-temperature RIXS
response disappear. Also, in the absence of a momentum selection
rule (e.g., $\mathbf{q} = \mathbf{k} + \mathbf{k}'$), a larger
number of processes become allowed and hence the energy range of the
RIXS response increases. Interestingly, both of these features are
already observable at $T = 0.1J \lesssim \Delta$ (see
Figs.~\ref{fig-2} and \ref{fig-3}).

\begin{figure}
\includegraphics[width=0.9\columnwidth]{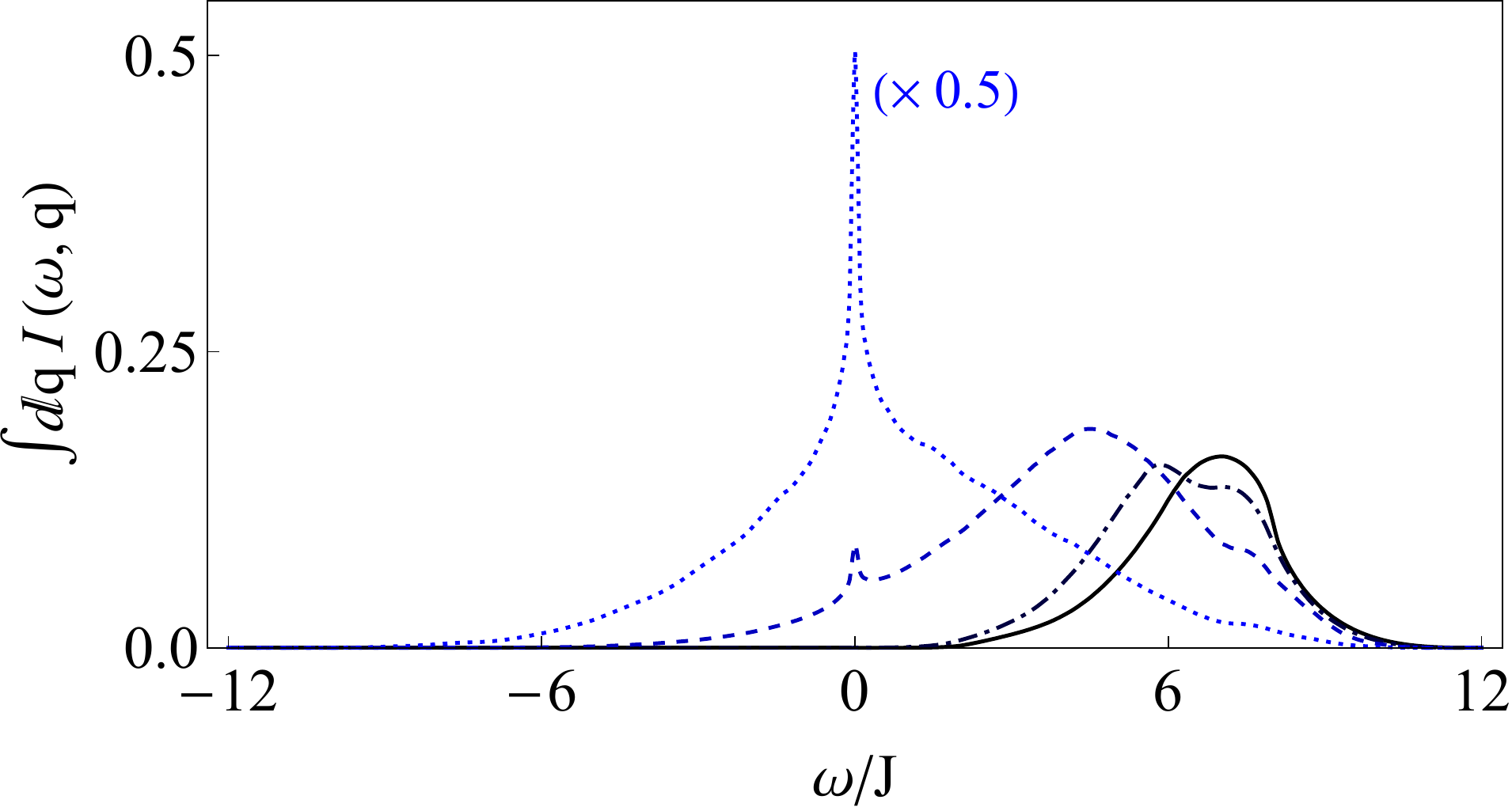}
\caption{Momentum-integrated RIXS intensity $I (\omega)$ for
temperatures $T = 0$ (solid line), $T = 0.1J$ (dashed-dotted line),
$T = J$ (dashed line), and $T = 5J$ (dotted line). The $T = 5J$
curve is multiplied by $0.5$ to compare with other
curves.}\label{fig-4}
\end{figure}

At temperatures $T \gtrsim T_H$, fermions become thermally excited
in large numbers, and Stokes processes are thus no longer dominant
over anti-Stokes and ``mixed'' processes. Consequently, the spectral
weight of the RIXS response is shifted to lower energies and becomes
nonzero even at $\omega < 0$. To distinguish this overall shift of
the spectral weight from the finer changes discussed in the previous
paragraph, we plot the momentum-integrated RIXS intensity $I
(\omega) \equiv \int d \mathbf{q} \, I (\omega, \mathbf{q})$ for a
range of different temperatures in Fig.~\ref{fig-4}. While the $T =
0.1J$ response is almost identical to the zero-temperature one, the
spectral weights of the $T \geq J$ responses are significantly
shifted to progressively smaller energies. In particular, the $T =
5J$ response is almost symmetric with respect to $\omega = 0$,
indicating that Stokes and anti-Stokes processes are almost equally
probable.

\begin{figure}
\includegraphics[width=0.99\columnwidth]{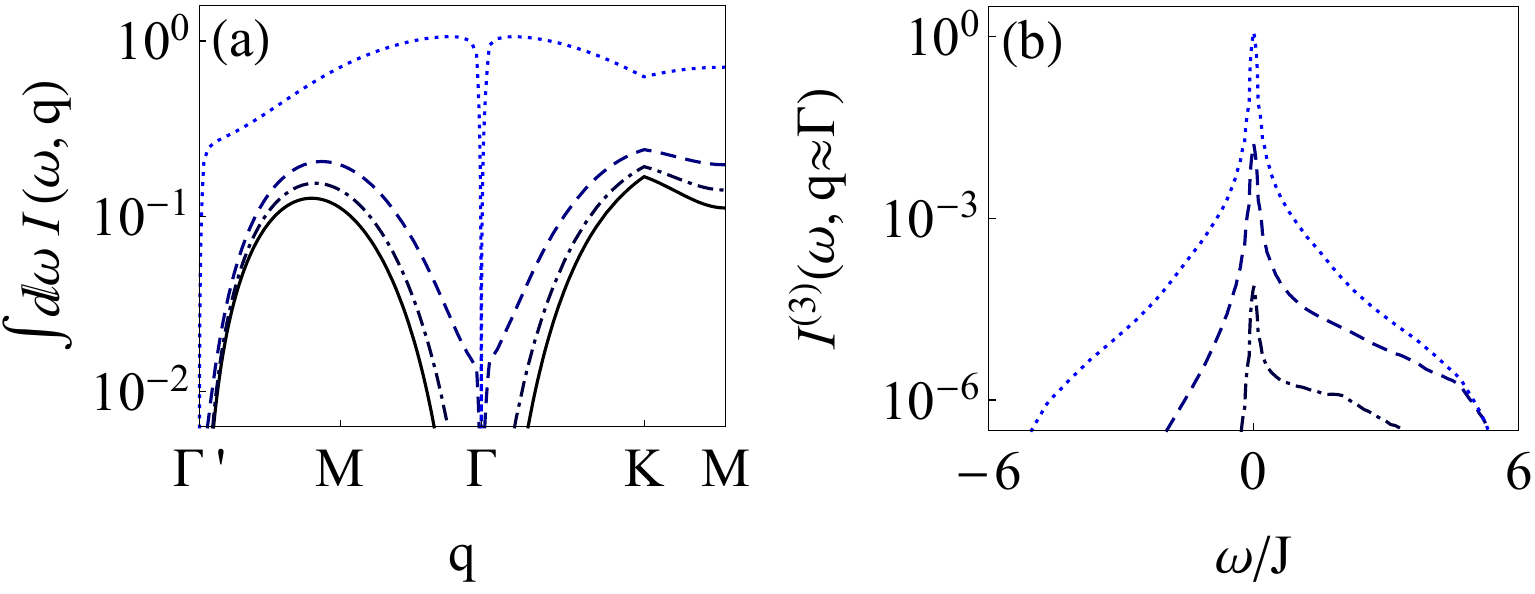}
\caption{(a) Energy-integrated RIXS intensity $I (\mathbf{q})$ along
the high-symmetry path $\Gamma^{\prime}$-M-$\Gamma$-K-M [see inset
of Fig.~\ref{fig-2}(a)] and (b) ``mixed'' RIXS intensity $I^{(3)}
(\omega, \mathbf{q})$ in a small region around the $\Gamma$ point.
In each subfigure, the curves correspond to temperatures $T = 0$
(solid line), $T = 0.1J$ (dashed-dotted line), $T = 0.5J$ (dashed
line), and $T = 5J$ (dotted line), and are normalized such that the
peak of the $T = 5J$ curve is at $1$. }\label{fig-5}
\end{figure}

Moreover, at temperatures $T \gg T_H$, the momentum-integrated RIXS
intensity exhibits a strong peak around zero energy as a result of
quasi-elastic ``mixed'' processes at small momenta $\mathbf{q}$.
These processes do not change the total number of fermions and
instead correspond to collective energy-density fluctuations
\cite{Brya1974, Richards1974, Reiter1976}. In Fig.~\ref{fig-5}, we
pinpoint the existence of these quasi-elastic processes in two
different ways. First, we plot the energy-integrated RIXS intensity
$I (\mathbf{q}) \equiv \int d \omega \, I (\omega, \mathbf{q})$ in
Fig.~\ref{fig-5}(a), and observe that its maximum is transferred
from the boundary to the center of the Brillouin zone upon
increasing the temperature. Second, we plot the ``mixed'' component
of the RIXS intensity, corresponding to the third term in
Eq.~(\ref{eq-I-2}), integrated over a small region around the
$\Gamma$ point of the Brillouin zone, in Fig.~\ref{fig-5}(b). This
quantity, $I^{(3)} (\omega, \mathbf{q} \approx \Gamma) \equiv
\int_{|\mathbf{q}| < \epsilon} d \mathbf{q} \, I^{(3)} (\omega,
\mathbf{q})$, is strongly peaked around zero energy, and its $\omega
= 0$ peak grows rapidly as the temperature is increased
\footnote{The integration \emph{around} the $\Gamma$ point is
required because the RIXS intensity vanishes at the $\Gamma$ point
itself.}. Interestingly, such a quasi-elastic peak has been
experimentally observed at high temperatures in the Raman response
of the Kitaev QSL candidate $\alpha$-RuCl$_3$ \cite{Sandilands2015,
Wang2018}.

We emphasize that the matrix elements $[\mathcal{A}_{\pm}
(\mathbf{q})]_{nn'}$ in Eq.~(\ref{eq-I-2}) also lead to observable
features in the RIXS response. First of all, the RIXS intensity $I
(\omega, \mathbf{q})$ vanishes for all energies $\omega$ at a
reciprocal lattice vector $\mathbf{q} = \mathbf{G}$ or,
equivalently, at the $\Gamma$ point of the Brillouin zone. Indeed,
since the RIXS vertex $R (\mathbf{G})$ in Eq.~(\ref{eq-R-2}) is
proportional to the Hamiltonian $H$ in Eq.~(\ref{eq-H-1}), it does
not create any excitations and gives rise to a purely elastic
response. Importantly, this connection between $H$ and $R
(\mathbf{G})$ is valid on the level of the spins, and the
corresponding suppression around the $\Gamma$ point is thus a robust
feature of the RIXS response at arbitrary temperature
\footnote{Note, however, that the complete extinction of the RIXS
intensity at the $\Gamma$ point seen in Figs.~\ref{fig-2} and
\ref{fig-5} may be partially obscured in RIXS experiments, possibly
appearing only as a suppression, due to the overlap with elastic
scattering intensity, finite momentum resolution, or both.}.

Furthermore, in comparison to the equivalent $\Gamma'$ points in the
neighboring Brillouin zones ($\mathbf{q} = \mathbf{G} \neq
\mathbf{0}$), the suppression of the RIXS response may be stronger
or weaker around the $\Gamma$ point in the central Brillouin zone
($\mathbf{q} = \mathbf{0}$) [see Fig.~\ref{fig-2}] due to a
destructive or constructive interference between RIXS processes at
the two sublattices of the bipartite honeycomb lattice. In general,
there is a complex phase factor $\pm i$ between the two sublattices
for each fermion created (annihilated). For the (anti-)Stokes
processes, dominating at low temperatures and/or far away from
$\omega = 0$, the interference at $\mathbf{q} = \mathbf{0}$ is
destructive due to $(\pm i)^2 = -1$, and the RIXS response is
\emph{weaker} around the central $\Gamma$ point. Conversely, for the
``mixed'' processes, dominating close to $\omega = 0$ at high
temperatures, the interference at $\mathbf{q} = \mathbf{0}$ is
constructive due to $(+i)(-i) = +1$, and the RIXS response is
\emph{stronger} around the central $\Gamma$ point. Importantly, the
phase factors $\pm i$ indicate that inversion symmetry acts
projectively on the fermions, and the stronger suppression around
the central $\Gamma$ point at low temperatures $T \lesssim T_L$ is
thus an indirect signature of their fractionalized nature
\cite{Halasz2016, Halasz2017}.

\section{Summary and outlook}

In this work, we presented a microscopic calculation of the
finite-temperature RIXS response for the Kitaev QSL on the honeycomb
lattice. In order to obtain a universal magnetic response, we
concentrated on indirect RIXS which only has one magnetic channel
and couples exclusively to the Majorana fermions. However, in stark
contrast to the case of zero temperature, thermally excited
$\mathbb{Z}_2$ fluxes are also indirectly observable at finite
temperature as they give rise to an effective disorder potential for
the Majorana fermions. In fact, as the temperature is increased, the
RIXS response changes qualitatively at two well-separated
temperature scales, $T_L$ and $T_H$, due to the thermal
proliferation of $\mathbb{Z}_2$ fluxes and Majorana fermions,
respectively. We thus conclude that the temperature evolution of the
RIXS response provides further evidence of spin fractionalization,
in addition to those already observable at zero temperature.

Moreover, as the small-momentum regime of RIXS is directly related
to Raman scattering, we can provide a possible explanation for the
strong quasi-elastic peak that has been experimentally observed in
the Raman response of $\alpha$-RuCl$_3$ \cite{Sandilands2015,
Wang2018}. Indeed, we found a similar quasi-elastic peak in our
theoretical RIXS response above the higher temperature scale $T_H$
and understood that it corresponds to long-wavelength collective
fluctuations of the Majorana fermions. While we were not able to
quantitatively reproduce its experimentally observed temperature
dependence, we argue that this strong quasi-elastic peak, which so
far has been subtracted as an unknown background, is also
qualitatively consistent with the presence of fractionalized
excitations in $\alpha$-RuCl$_3$.

Finally, we emphasize that our results capture the fundamental
properties of the indirect RIXS process at the $K$-edge of Ru$^{3+}$
in $\alpha$-RuCl$_3$. The predicted energy resolution at this edge,
$\Delta \omega \sim 1$ meV \cite{Gog2013, Gog2016}, is much smaller
than the bandwidth $12J \sim 20$ meV \cite{Winter2016} of the
magnetic RIXS response. Due to this favorable prediction for the
energy resolution and the universality of the corresponding RIXS
response, we hope that, in the near future, the quantitative
predictions in this work will serve as a useful guide for RIXS
experiments in Kitaev materials.

\begin{acknowledgments}

We are grateful to K.~Burch and J.~H.~Kim for valuable discussions.
The work of G.~B.~H. was supported at ORNL by Laboratory Director's
Research and Development funds and at the KITP by the Gordon and
Betty Moore Foundation's EPiQS Initiative through Grant
No.~GBMF4304. S.~K. was supported in part through the Boston
University Center for Non-Equilibrium Systems and Computation.
N.~B.~P. was supported by the U.S. Department of Energy, Office of
Science, Basic Energy Sciences under Award No.~DE-SC0018056.
N.~B.~P. also acknowledges the hospitality of the KITP and the NSF
Grant No.~PHY-1748958, where the work was initiated.

\end{acknowledgments}

\bibliography{Kitaev_RIXS_finite_T_refs}
\end{document}